\documentclass[article]{aa}

\usepackage{graphicx}
\usepackage{txfonts}
\usepackage[colorlinks=true, citecolor=blue, linkcolor=blue, urlcolor=blue]{hyperref}
\begin{document}

%%%%%%%%%%%%%%%%%%%%%%%%%%%%%%%%%%%%%%%%
% if you use custom commands in your title,
% ensure to check your title when submitting!
%%%%%%%%%%%%%%%%%%%%%%%%%%%%%%%%%%%%%%%%
    \title{Spectral Hardening Revealed by Geometric De-boosting in the Masked Jet of PKS 2155$-$304}

   %\subtitle{Subtitle}

    \titlerunning{Spectral Hardening via Geometric De-boosting in PKS 2155-304}

%%%%%%%%%%%%%%%%%%%%%%%%%%%%%%%%%%%%%%%%
% Please separate each author with the \and command
%
% Please do not include ORCIDs next to author names.
% Only ORCIDs authenticated by individual authors in EDPS
% editorial system will be taken into account.
% ORCIDs included here will be removed.
%%%%%%%%%%%%%%%%%%%%%%%%%%%%%%%%%%%%%%%%

\author{Alberto Dom\'inguez,\inst{1,2}\fnmsep\thanks{\email{alberto.d@ucm.es}}
       Adithiya Dinesh,\inst{1,2}%\fnmsep\thanks{A. Dinesh and E. Madero contributed equally to this work.}
       \and Elena Madero\inst{1}%\fnmsep\thanks{A. Dinesh and E. Madero contributed equally to this work.}
       }

\institute{$^1$Department of EMFTEL, Universidad Complutense de Madrid, E-28040 Madrid, Spain \\
            $^2$Instituto de Física de Partículas y del Cosmos (IPARCOS), Universidad Complutense de Madrid, E-28040 Madrid, Spain
            }

   \date{Received February 24, 2026}

% \abstract{}{}{}{}{}
% 5 {} token are mandatory
 
\abstract
% context heading (optional)
{Blazar gamma-ray variability is predominantly stochastic and well described by red-noise processes. However, a subset of sources shows quasi-periodic oscillations (QPOs) on year-long timescales, whose physical origin, geometric or plasma-driven, remains debated. In high-synchrotron-peaked (HSP) blazars, departures from a single power-law gamma-ray spectrum, manifested as high-energy upturns in the GeV band, may probe emission mechanisms operating in these sources and their intrinsic duty cycle.}
% aims heading (mandatory)
{We investigate the link between the 1.7\,yr gamma-ray QPO in PKS~2155$-$304 and an exceptional spectral hardening event identified in the \textit{Fermi}-LAT population. By mapping this spectral state onto the periodic cycle, we aim to constrain the mechanism driving the oscillation and conditions under which plasma-driven acceleration signatures become visible.}
% methods heading (mandatory)
{We analyze first 17.4 years of \textit{Fermi}-LAT data using 30-day binning. To isolate variability on QPO timescales, we apply Singular Spectrum Analysis to mitigate red-noise effects and use a Moving Block Bootstrap approach to quantify correlations between photon flux and photon index while accounting for temporal autocorrelation.}
% results heading (mandatory)
{We find a statistically significant softer-when-brighter chromatic trend, atypical for HSP blazars and supporting a geometric origin for the flux modulation. Crucially, the spectral hardening event is phase-locked to the QPO trough. This anti-correlation implies that the hardening signature is detectable only when geometrically boosted soft emission is sufficiently suppressed at the flux minimum.}
% conclusions heading (optional)
{We propose a Geometric Masking scenario in which relativistic jet geometry regulates the visibility of microphysical acceleration processes. The detection of spectral hardening at the flux minimum suggests that such events may be a universal but masked feature of blazar jets, revealed only when Doppler boosting is minimized. These results favor a two-component jet structure and imply that searching for spectral hardening during low-flux states, even in non-periodic sources, may be essential for unveiling the underlying jet physics obscured by relativistic amplification.}


   \maketitle
\nolinenumbers

%%%%%%%%%%%%%%%%%%%%%%%%%%%%%%%%%%%%%%%%%%%%%%%%%%%%%%%%%%%%%%
\section{Introduction}
The variability of blazars in the gamma-ray band is predominantly stochastic, typically described by power-law noise densities \citep[e.g.,][]{Abdo2010, Vaughan2016}. Against this chaotic background, a small subset of sources shows compelling evidence for quasi-periodic oscillations (QPOs) on year-long timescales \citep[e.g.,][]{Ackermann2015, Penil2020, Ren2023, Rico2025}. The physical origin of these rare periodicities remains a subject of debate, with proposed mechanisms broadly divided into intrinsic plasma-driven processes, such as recurrent magnetic reconnection or quasi-periodic shocks \citep[e.g.,][]{Zhang2014, Shukla2018}, and geometric origins, where the modulation arises from spatial reorientation, maybe driven by supermassive binary black hole systems that induce periodic jet precession. \citep[e.g.,][]{Begelman1980, Rieger2004, Caproni2017}.

While light-curve flux analyses alone generally fail to distinguish between these scenarios, phase-resolved spectral analysis provides a critical diagnostic path \citep{Madero2026}. In standard shock-in-jet models, flux increases are driven by particle injection or acceleration, typically leading to a chromatic harder-when-brighter spectral evolution \citep[e.g.,][]{Kirk1998}. In contrast, geometric models attribute the modulation to periodic variations in the Doppler factor ($\delta$). Since relativistic beaming amplifies the bolometric flux ($F \propto \delta^p$) without necessarily altering the intrinsic electron energy distribution, geometric drivers are expected to produce variability that is largely achromatic \citep{Madero2026}. However, if the boosted emission is dominated by cooled downstream populations or a softer jet sheath, geometric viewing angle changes can result in a softer-when-brighter trend \citep[e.g.,][]{Raiteri2017}, a signature clearly distinct from shock-driven acceleration.

The high-synchrotron-peaked (HSP) blazar PKS~2155$-$304 offers a unique opportunity to test these predictions. Long-term monitoring with the \textit{Fermi} Large Area Telescope (LAT) has established it as one of the most robust QPO candidates, with a characteristic period of $P \approx 1.7$\,yr reported by multiple independent studies \citep[e.g.,][]{Sandrinelli2014, Zhang2017, Penil2020, Rico2025}. Recently, a systematic search for extreme spectral variability in the \textit{Fermi}-LAT population identified PKS~2155$-$304 as one of only two sources showing a statistically significant spectral hardening event during a flaring episode \citep{Dinesh2025}. This rare event, where the spectrum deviated sharply from a power law, raises a critical question: is this anomaly a random stochastic fluctuation, or is its visibility regulated by the source's periodic modulation?

In this Letter, we investigate the physical link between the 1.7\,yr QPO and the transient spectral hardening in PKS~2155$-$304, defining spectral hardening as a departure from a single power-law gamma-ray spectrum, manifested as high-energy upturns in the GeV band. Using 17.4 years of \textit{Fermi}-LAT data, we characterize the flux vs. photon index behavior and map the hardening event onto the phase of the oscillation. We find that the bulk variability is atypical for an HSP, showing a softer-when-brighter trend with no significant correlation between photon index and phase, consistent with the coexistence of stochastic plasma-driven activity and a geometric modulation. In addition, the uncommon spectral hardening event is phase-locked to the QPO trough.

We interpret these findings within a Geometric Masking scenario, in which the periodic minimization of the Doppler factor suppresses the dominant soft background and reveals intrinsic acceleration signatures that remain masked during high-flux states. In this picture, the high-flux baseline reflects maximum geometric masking rather than intrinsic quiescence.

%%%%%%%%%%%%%%%%%%%%%%%%%%%%%%%%%%%%%%%%%%%%%%%%%%%%%%%%%%%%%%
\section{Data and Methodology}\label{sec:data}

We analyzed the $\gamma$-ray emission of PKS~2155$-$304 using approximately 17.4 years of \textit{Fermi}-LAT data (August 2008--January 2026; MJD 54698--61058) obtained from the \textit{Fermi}-LAT Light Curve Repository\footnote{\url{https://fermi.gsfc.nasa.gov/ssc/data/access/lat/LightCurveRepository/}} \citep[LCR;][]{Abdollahi2023}. We utilized the standard LCR light curves integrated strictly over the 100~MeV to 100~GeV energy range. As a well-established TeV blazar with a characteristically hard \textit{Fermi}-LAT spectrum (average photon index $\Gamma \approx 1.8$), the observed $\gamma$-ray flux of PKS~2155$-$304 is driven by high-energy photons \citep[e.g.,][]{Aharonian2007, Ackermann2016}. To characterize variability on the timescales of the reported QPO, we adopted a 30-day binning with a minimum test statistic threshold of $TS \ge 4$ per bin and allowed the photon index ($\Gamma$) to vary freely in each bin.

To isolate variability on the QPO timescale and reduce red-noise contamination, we applied Singular Spectrum Analysis (SSA), a non-parametric time-series decomposition technique that separates long-term trends, oscillatory components, and stochastic noise. We reconstructed the baseline trend and used the detrended light curve for the subsequent correlation analyses. In addition, we reconstructed the QPO component itself \citep[for details on the SSA setup and configuration, see][]{Rico2025}.

The spectral behavior was then mapped onto our reconstructed periodic modulation. Using the Lomb-Scargle method \citep{VanderPlas2018}, we obtained a clear peak corresponding to a period of $P = 1.7~\mathrm{yr}$ ($\approx 621,\mathrm{days}$), in agreement with values reported in recent long-term studies \citep{Zhang2017, Penil2020, Rico2025}. We do not formally compute the significance in this work; however, \citet{Rico2025}, who use the same methodology for the QPO characterization, report a local significance of $4.5\sigma$ and $3\sigma$ after trial corrections, representing strong evidence for a year-scale QPO in the context of red-noise-dominated blazar variability. Crucially, we defined the reference phase $\phi = 0$ at the epoch of the QPO flux maximum occurring prior to the spectral hardening event (MJD 57398). This choice anchors the phase to the state of maximum geometric boosting, assuming a Doppler origin for the QPO, and allows us to interpret phase-dependent spectral changes relative to the boosted state.

The correlation between flux and photon index was quantified using the Spearman rank coefficient ($\rho$). This non-parametric approach is particularly well suited for the analysis of PKS~2155$-$304 because, unlike linear correlation coefficients, $\rho$ evaluates the strength of a monotonic relationship between variables. This is especially appropriate for blazar spectral evolution, where the coupling between flux ($F$) and photon index ($\Gamma$) frequently follows complex, non-linear power-law or exponential trends rather than a strictly linear dependence. Uncertainties were estimated using two independent bootstrap methods: a Standard Bootstrap (SB) and a Moving Block Bootstrap (MBB). Following \cite{Madero2026}, we note that while the MBB preserves the temporal structure of the red noise, it may introduce an attenuation bias that slightly underestimates the correlation strength. We therefore employ both the SB (which assumes independence) and the MBB (block size $k=4$, or 120 days) to bracket the true uncertainty and assess the robustness of the results, with sensitivity tests performed across $k=2$ (60 days) to $k=6$ (180 days). As described in the next section, both methods yield consistent results. Finally, we overlaid the window of the spectral hardening event identified by \cite{Dinesh2025} onto the phase-folded light curve to determine its alignment with the periodic cycle.

%%%%%%%%%%%%%%%%%%%%%%%%%%%%%%%%%%%%%%%%%%%%%%%%%%%%%%%%%%%%%%
\section{Results}\label{sec:results}

\subsection{Global Spectral Behavior}

\begin{figure}
\centering
\includegraphics[width=\columnwidth]{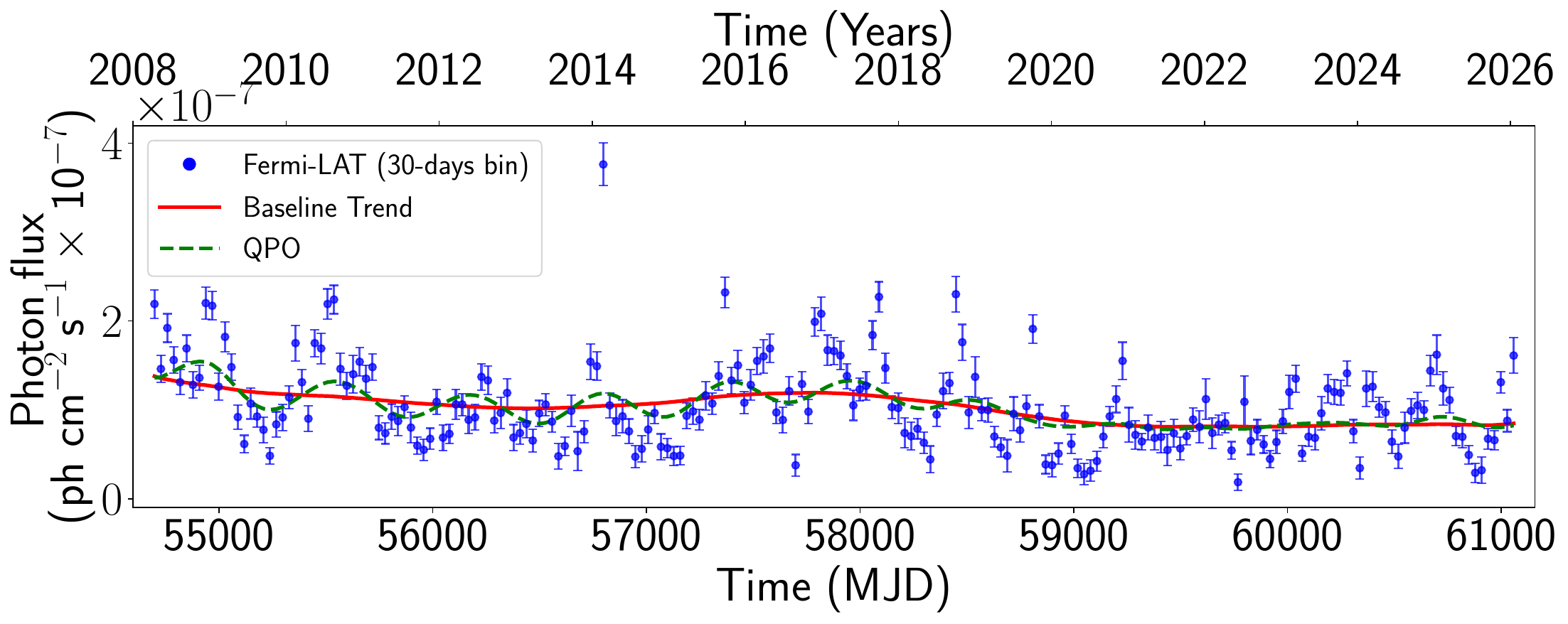}
\caption{Gamma-ray light curve of PKS~2155$-$304 (30-day bins, 2008--2026): Fermi-LAT flux (blue circles, $1\sigma$ errors), the SSA baseline (red line) used for detrending, and the reconstructed QPO component (green dashed line).}
\label{Fig:LightCurve}
\end{figure}

\begin{figure*}[ht]
    \centering
    \includegraphics[width=0.45\textwidth]{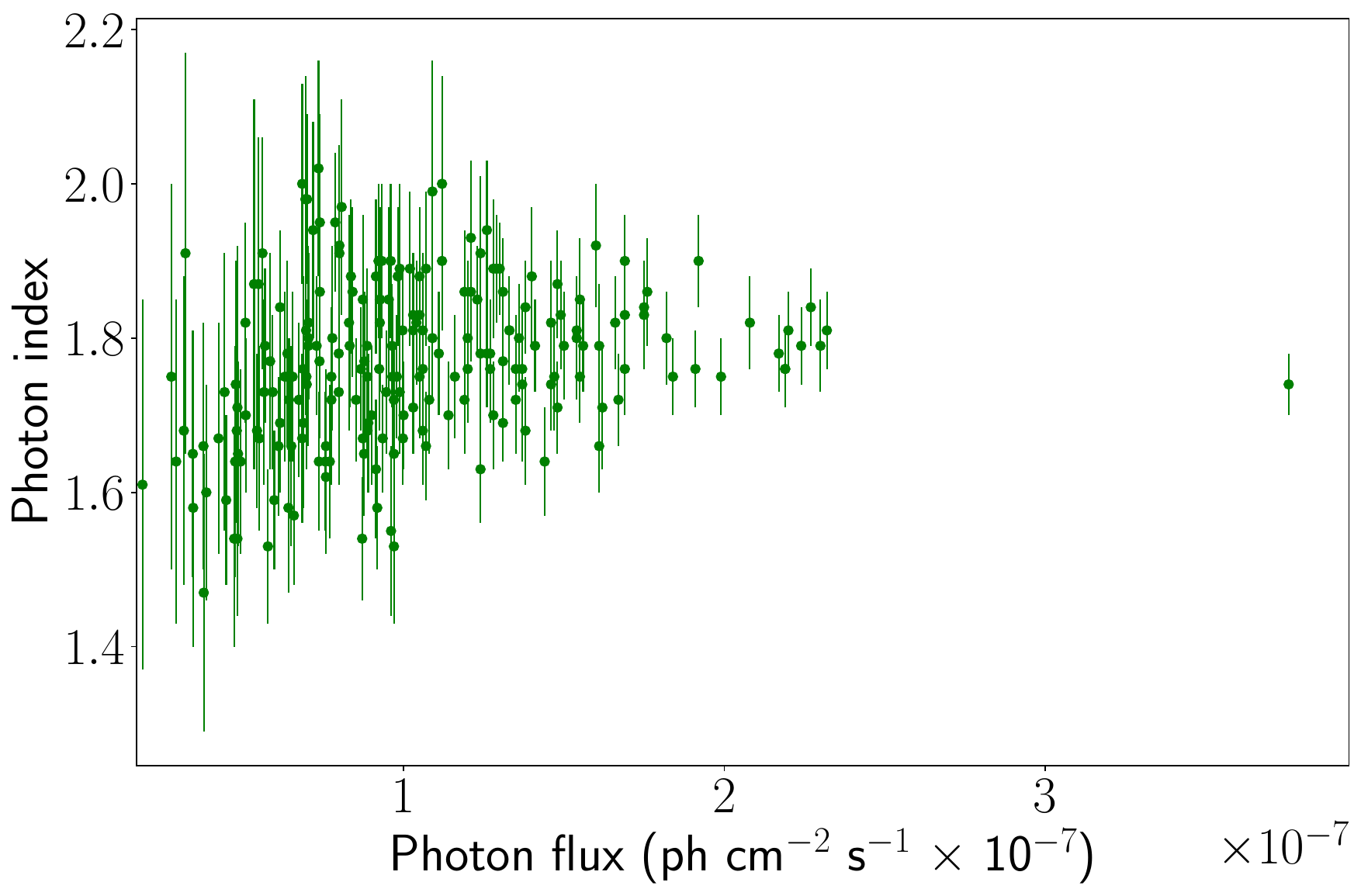}
    %\hfill
    \includegraphics[width=0.45\textwidth]{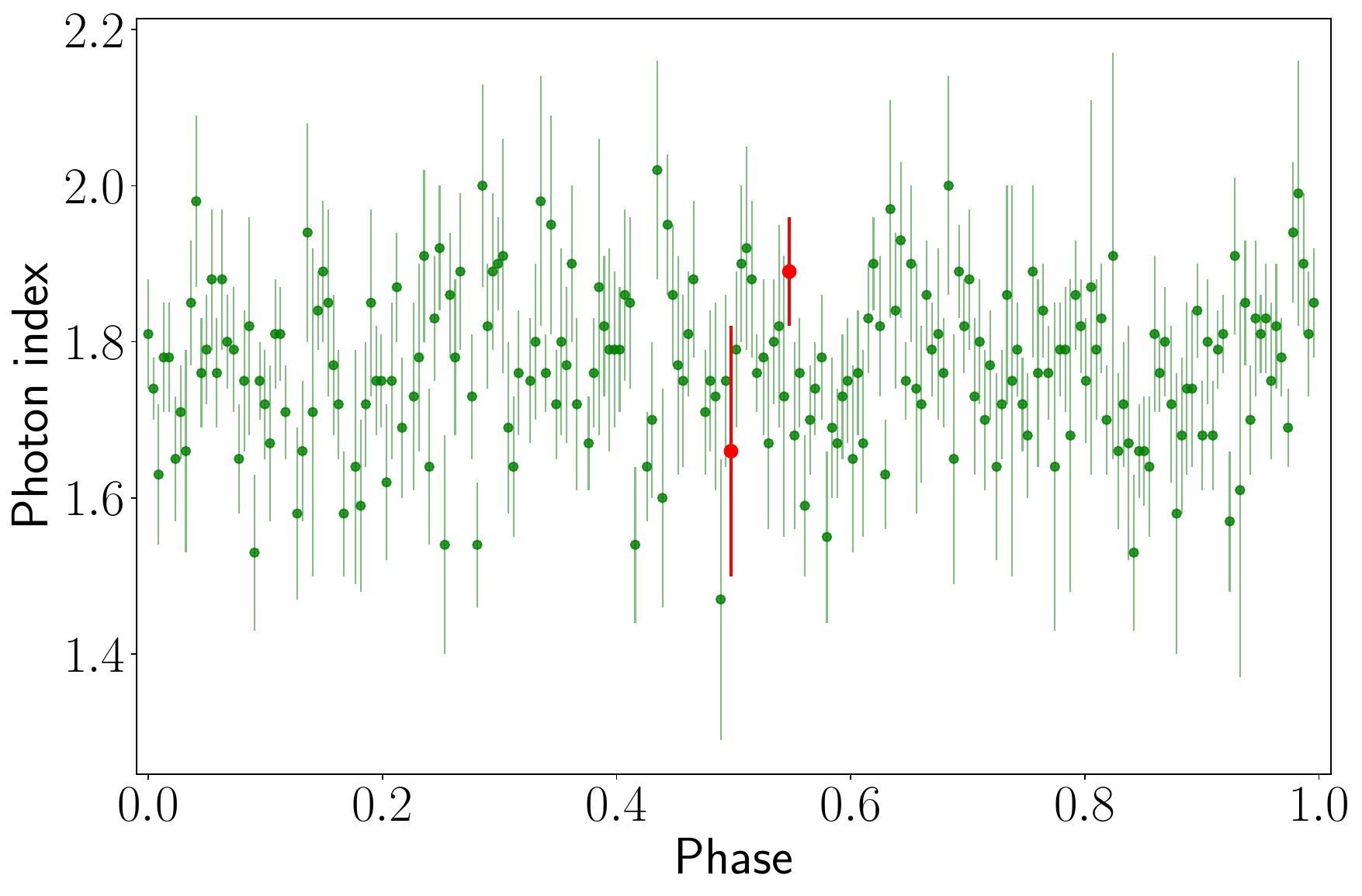}
    \caption{Spectral variability of PKS~2155$-$304. \textit{Left:} Photon index vs. photon flux, showing a positive correlation ($\rho \approx 0.3$), weak but robust, indicative of an atypical softer-when-brighter trend. \textit{Right:} Phase-resolved spectral evolution folded at the 1.7\,yr period, with the 17.4-year baseline variability (green dots) consistent with an achromatic trend given the lack of significant correlation between photon index and phase ($\rho = 0.036$, $p = 0.60$), and the two consecutive 30-day bins of the extreme spectral hardening event reported by \citet[][red dots]{Dinesh2025}.}
    \label{fig:spectral_combined}
\end{figure*}

We first characterized the global spectral variability of PKS~2155$-$304 over the full 17.4-year baseline to establish a physical reference for the QPO modulation (Figure \ref{Fig:LightCurve}). Figure~\ref{fig:spectral_combined} illustrates the relationship between the photon flux and the photon index. In contrast to the harder-when-brighter trend typical of most HSP blazars, PKS~2155$-$304 shows a statistically significant positive correlation. 

Using the detrended data obtained after reconstructing and subtracting the baseline trend with SSA, we mitigated red-noise contamination and measured a Spearman rank coefficient of $\rho = 0.289$ ($p < 2 \times 10^{-5}$). This result is consistent with the original correlation from the observed light curve ($\rho = 0.318$, $p < 3 \times 10^{-6}$). It remains robust when accounting for statistical uncertainties. The Standard Bootstrap estimates an uncertainty of $\pm 0.063$, while the red-noise-accounting Moving Block Bootstrap ($k=4$) yields a consistent uncertainty of $\pm 0.064$. Sensitivity tests across block sizes $k=2$ to $k=6$ confirm this stability. This robust softer-when-brighter behavior implies that the highest flux states are not driven by the injection of fresh, high-energy particles (which would harden the spectrum). 

Furthermore, we detected no significant correlation between the photon index and the QPO phase itself ($\rho = 0.036, p=0.60$, indicating that the periodic modulation is effectively achromatic on average. This decoupling—a global softer, when-brighter trend in the flux-index plane but an achromatic QPO phase profile, mirrors the behavior reported for PG~1553+113 \citep{Madero2026}. It suggests that two distinct mechanisms coexist in the jet: a geometric driver (e.g., precession) responsible for the periodic, achromatic/soft baseline modulation, and stochastic plasma-driven processes responsible for the superimposed flaring activity.

\subsection{Phase-Locked Spectral Hardening}

Against this global background of soft/achromatic variability, the extreme spectral hardening event (GeV-band upturn deviating from power-law) identified by \cite[][MJD 57692--57748; $|\Delta\Gamma| \approx 2.54\pm 0.32$, $TS_{hard}=16$]{Dinesh2025} stands out as an anomaly beyond $3.5\sigma$. As shown in Figure~\ref{fig:spectral_combined}, mapping this event onto the periodic 1.7\,yr cycle reveals a striking phase-locking phenomenon: the hardening event occurs at a phase of $\phi \approx 0.53$, coinciding with the trough of the quasi-periodic modulation.

The spectral hardening episode (MJD 57692--57748) is resolved into two consecutive 30-day bins in our analysis. In the phase-folded representation (Fig.~\ref{fig:spectral_combined}), these bins appear separated by the characteristic phase step of $\Delta \phi \approx 0.05$, corresponding to the ratio of the 30-day bin width to the 1.7\,yr period. Their strict clustering around the QPO trough at $\phi \approx 0.53$ confirms that the hardening event is phase-locked to the flux minimum.

To quantify the significance of this alignment, we performed a phase-scrambling Monte Carlo test ($10^{4}$ realizations) in which the MJD window of the hardening event was randomly reassigned to any valid 30-day bin within the 17.4-year baseline, and the resulting phase was recorded. The probability of a random placement falling within $|\Delta \phi| \leq 0.03$ of the trough center is $p \approx 0.059$. Combining this with the independent flux--index correlation result ($p \approx 1.8 \times 10^{-5}$) under the assumption of statistical independence, the joint probability of both features arising by chance is $p_{\mathrm{joint}} \approx 1.1 \times 10^{-6}$, strongly disfavoring a coincidental alignment. We interpret the physical origin of this phase-locked behavior in the following section. Note that the assumption of independence is reasonable here, as the global flux-index correlation reflects the overall chromatic behavior of the source across the full 17.4-year dataset, while the phase alignment test specifically evaluates the timing of a single transient hardening event relative to the periodic cycle. These probe distinct physical and statistical aspects of the variability, with no evident coupling in the underlying data generation process.

%%%%%%%%%%%%%%%%%%%%%%%%%%%%%%%%%%%%%%%%%%%%%%%%%%%%%%%%%%%%%%
\section{Discussion}\label{sec:discussion}

\subsection{The Geometric Masking Scenario}

The phase alignment of the spectral hardening event with the QPO trough ($\phi \approx 0.53$) suggests a conditional visibility mechanism. This trend is naturally explained by geometric Doppler-modulation models, a foundational framework in blazar physics \citep[e.g.,][]{Blandford1978, Urry1995, Rieger2004}. Relativistic beaming amplifies flux according to $F \propto \delta^{3+\alpha}$, where $\alpha$ is the spectral index ($F_\nu \propto \nu^{-\alpha}$) \citep[e.g.,][]{Lind1985, Dermer1995}. Because the soft emission envelope (e.g., a cooled electron population or jet sheath) has a steeper index than the stochastic hard core, it experiences a stronger differential Doppler boost \citep[e.g.,][]{Ghisellini2005}. Consequently, at the flux maximum (smallest viewing angle), the geometrically boosted soft emission overwhelms the hard core, naturally producing the observed softer-when-brighter trend \citep[e.g.,][]{Villata2004, Raiteri2017}. The presence of such active Doppler factor variations in PKS~2155$-$304 has been independently confirmed by contemporary multi-epoch SED modeling \citep{Harutyunyan2026}. We emphasize that this is a phenomenological application of the same geometric framework recently utilized to interpret the QPO in PG~1553+113 \citep{Madero2026}.

In this Geometric Masking scenario, we propose that the observational rarity of spectral hardening in HSP blazars ($<0.1\%$; \citealt{Dinesh2025}) is not necessarily an indication that extreme particle acceleration events are uncommon, but rather that they are frequently masked. In a geometric QPO scenario, flux peaks are driven by maximum relativistic beaming. Our finding of a global softer-when-brighter trend ($\rho \approx 0.3$) indicates that these boosted states are dominated by a soft photon field, likely originating from cooled electron populations or the jet sheath. This may redefine the high-flux state not as a period of intrinsic quiescence, but as a regime of maximum geometric masking. The jet may host frequent plasma-driven particle acceleration events that remain observationally unresolved, overwhelmed by the intensified soft background. During these phases, the spectral signature of intrinsic acceleration, such as shocks or magnetic reconnection, is suppressed by the boosted soft emission. Conversely, at the QPO trough, the Doppler factor is minimized, effectively suppressing the geometric boost and reducing this soft background. This reduces the geometric masking, allowing hard-spectrum emission from transient jet instabilities to dominate the observed flux and making the hardening detectable.

This mechanism implies a significant selection bias in current blazar surveys. If hardening events occurring during flux peaks remain observationally diluted, the intrinsic duty cycle of extreme particle acceleration in HSPs may be substantially higher than previously estimated.

An important corollary of this scenario addresses the rarity of the detected event. While the masking effect is minimized periodically every 1.7\,yr, \cite{Dinesh2025} reported only one significant spectral hardening episode over the 17.4-year baseline. This apparent scarcity is likely due to two factors. First, the geometric minimum is a necessary condition for visibility, not a driver of acceleration; an intrinsic stochastic shock must still coincide with the narrow temporal window of the trough. Second, the blind search strategy employed by \cite{Dinesh2025} relied on high-significance deviations which favor high-flux states. Since QPO troughs are by definition low-flux intervals, weaker or shorter-duration hardening events may have failed to meet the strict global detection thresholds. We predict that a targeted search specifically focused on the QPO troughs of PKS~2155$-$304 will reveal a population of sub-threshold hardening events that were missed by previous blind surveys.

\subsection{Constraints on Jet Structure and Extreme States}

The coexistence of a stable geometric modulation and transient spectral hardening provides a unique constraint on the jet's spatial architecture. While purely plasma-driven QPO models, such as recurrent shocks \citep[e.g.,][]{Kirk1998}, could in principle produce similar periodicities, they generally predict harder-when-brighter spectral evolution, which is inconsistent with our observed softer-when-brighter trend and achromatic phase profile. This favors the geometric interpretation, where the achromatic nature of the global QPO suggests it arises from large-scale geometric changes affecting a steady emission envelope. In contrast, the phase-locked hardening represents localized microphysical processes. This separation of scales implies a two-component jet structure: a soft-spectrum envelope modulated by geometry and a stochastic core where intermittent hard shocks occur. Recent time-resolved SED modeling of PKS 2155-304 over the same epoch supports this duality, revealing that while many flaring states are driven by bulk Doppler factor variations, isolated gamma-ray flares require localized hardening of the injected electron distribution without bulk geometric changes \citep{Harutyunyan2026}.

The physical uniqueness of this event is underscored by its departure from the source's typical spectral behavior. While high-flux states, including the decade-scale peak ($F \approx 3.8 \times 10^{-7}$\,ph\,cm$^{-2}$\,s$^{-1}$), are well-described by standard power-law models with indices consistent with the global mean, the trough-locked event shows a significant spectral break. This indicates that the event is not merely a hard fluctuation within a power-law distribution, but a distinct physical state characterized by a different emission mechanism or particle distribution. Its appearance only at the QPO trough reinforces the conclusion that such complex spectral states are intrinsically masked by the geometrically-boosted power-law background during flux peaks. Physically, this scenario is consistent with an emission region moving along a helical trajectory \citep[e.g.,][]{Sobacchi2017}. The global modulation arises from changing viewing angles, while the phase-locked hardening may reflect the component encountering a standing shock \citep{Marscher2008} precisely at the Doppler minimum, allowing the hard emission to dominate the suppressed baseline.

\subsection{Comparison with accretion-shock resonance models}
Alternative models, such as accretion-shock resonance oscillations \citep[e.g.,][]{Molteni1996, Dash2026}, can produce phase-locked spectral changes via coronal thermal Comptonization in radio-quiet active galactic nuclei and X-ray binaries. However, PKS~2155$-$304 is a highly beamed blazar whose $>100$\,MeV gamma-ray emission is generally attributed to non-thermal Synchrotron Self-Compton (SSC) processes within the relativistic jet \citep[e.g.,][]{Costamante2002, Katarzynski2008, Kusunose2008, Aharonian2009}. Recent broadband SED modeling over the 2008--2023 timeframe finds that the emission is well-described by a leptonic SSC jet model, making accretion disk thermal Comptonization a less likely primary driver for the gamma-ray flux \citep{Harutyunyan2026}. Furthermore, accretion-shock models typically predict a continuous, systematic phase-dependent spectral evolution (e.g., persistently hardest at phase $\sim 0.5$). The overall achromatic QPO phase profile in our data ($\rho = 0.036$), punctuated by a single rare hardening event, is less consistent with a persistent disk-shock driver and aligns more closely with geometric Doppler variations.

\subsection{Comparison with PG~1553+113 and Population Context}
The spectral phenomenology of PKS~2155$-$304 shares same large-scale geometric modulation as PG~1553+113, yet extends the physical picture through the detection of phase-dependent masking. Both sources host robust year-scale QPOs, $P \approx 1.7,\mathrm{yr}$ and $P \approx 2.2,\mathrm{yr}$ respectively \citep{Ackermann2015,Penil2020,Abdollahi2024,Penil2024,Rico2025}, and both show a softer-when-brighter flux--index trend that deviates from the standard HSP paradigm. We propose that this combination defines a spectral diagnostic for a distinct sub-class of geometrically modulated blazars, characterized by three necessary signatures: (i) a stable year-scale QPO; (ii) a global achromatic or softer-when-brighter chromatic trend ($\rho > 0$ in photon index versus flux space), inconsistent with shock-driven particle injection; and (iii) rare, trough-locked spectral hardening events.

PKS~2155$-$304 satisfies all three criteria. For PG~1553+113, criteria (i) and (ii) are already established in the literature. Whether criterion (iii) is also met, that is, whether spectral hardening events are present and phase-locked to its QPO trough, remains an open observational question and a direct prediction of our framework. A systematic phase-resolved spectral analysis of PG~1553+113 analogous to the approach of \citet{Dinesh2025} would provide a direct test of the Geometric Masking scenario and of the broader hypothesis that this sub-class is physically distinct from stochastic HSP blazars.

In this picture, the large-scale jet geometry acts as a stable regulator of the visibility of microphysical processes, distinguishing these sources from purely stochastic blazars, where variability is dominated by the harder-when-brighter signature of particle injection. If geometric masking is a generic feature of Doppler-modulated HSPs, PG~1553+113 represents the most immediate and accessible target to confirm or falsify this interpretation.

%%%%%%%%%%%%%%%%%%%%%%%%%%%%%%%%%%%%%%%%%%%%%%%%%%%%%%%%%%%%%%

\section{Summary and Conclusions}\label{sec:conclusions}

By combining the spectral search of \cite{Dinesh2025} with the analysis of the year-scale periodic modulation of PKS~2155$-$304 over 17.4 years of observations, we identify a physical connection between jet geometry and transient particle acceleration. Our main conclusions are as follows: (1) PKS~2155$-$304 shows a softer-when-brighter behavior supporting a geometric origin for its 1.7\,yr oscillation. (2) The exceptional spectral hardening event reported in \cite{Dinesh2025} is phase-locked to the QPO trough, coinciding with the minimum of the flux modulation. (3) This alignment supports a Geometric Masking scenario, where the suppression of boosted soft emission at the flux minimum enables the detection of otherwise masked acceleration signatures.

Beyond periodic systems, these results suggest a broader paradigm for blazar variability. We propose that the observational rarity of extreme spectral states across the entire blazar population may be a consequence of geometric masking. In this framework, many ``stochastic'' blazars may host frequent plasma-driven acceleration events that remain hidden by persistent relativistic boosting of a softer jet component. If this is the case, spectral hardening is not an intrinsic rarity of jet physics, but an observational signature that is filtered by the observer's viewing angle and the jet's Lorentz factor. Future high-sensitivity monitoring should prioritize spectral analysis during low-flux states, even in non-periodic sources, to unveil the underlying jet physics typically obscured by geometric amplification.

%%%%%%%%%%%%%%%%%%%%%%%%%%%%%%%%%%%%%%%%%%%%%%%%%%%%%%%%%%%%%%
%\noindent
\section*{Data availability}
Data are available in the \textit{Fermi}-LAT Light Curve Repository.

%%%%%%%%%%%%%%%%%%%%%%%%%%%%%%%%%%%%%%%%%%%%%%%%%%%%%%%%%%%%%%
\begin{acknowledgements}
This work has made use of public \textit{Fermi} data obtained from the High Energy Astrophysics Science Archive Research Center (HEASARC), provided by NASA Goddard Space Flight Center. We specifically acknowledge the \textit{Fermi} Science Support Center (FSSC) for the provision of the \textit{Fermi}-LAT Light Curve Repository. 
\end{acknowledgements}

%%%%%%%%%%%%%%%%%%%%%%%%%%%%%%%%%%%%%%%%%%%%%%%%%%%%%%%%%%%%%%
% WARNING
% Please note that we have included the references below in
% order to compile the document, but we ask you to:
%
% - use BibTeX with the regular commands:
%   \bibliographystyle{aa} % style aa.bst
%   \bibliography{Yourfile} % your references Yourfile.bib

\begin{thebibliography}{}

\bibitem[Abdo et al.(2010)]{Abdo2010}
Abdo, A.~A., Ackermann, M., Ajello, M., et al. 2010,
ApJ, 710, 1271

\bibitem[Abdollahi et al.(2023)]{Abdollahi2023}
Abdollahi, S., Ajello, M., Baldini, L., et al. 2023,
ApJS, 265, 31

\bibitem[Abdollahi et al.(2024)]{Abdollahi2024}
Abdollahi, S., Baldini, L., Barbiellini, G., et al. 2024,
ApJ, 976, 203

\bibitem[Ackermann et al.(2015)]{Ackermann2015}
Ackermann, M., Ajello, M., Albert, A., et al. 2015,
ApJ, 813, L41

\bibitem[Ackermann et al.(2016)]{Ackermann2016}
Ackermann, M., Ajello, M., Atwood, W.~B., et al. 2016,
ApJS, 222, 5

\bibitem[Aharonian et al.(2007)]{Aharonian2007}
Aharonian, F., Akhperjanian, A.~R., Bazer-Bachi, A.~R., et al. 2007,
ApJ, 664, L71

\bibitem[Aharonian et al.(2009)]{Aharonian2009}
Aharonian, F., Akhperjanian, A.~G., Anton, G., et al. 2009,
ApJ, 696, L150

\bibitem[Aleksi{\'c} et al.(2012)]{Aleksic2012}
Aleksi{\'c}, J., Alvarez, E.~A., Antonelli, L.~A., et al. 2012,
A\&A, 544, A75

\bibitem[Begelman et al.(1980)]{Begelman1980}
Begelman, M.~C., Blandford, R.~D., \& Rees, M.~J. 1980,
Nature, 287, 307

\bibitem[Blandford \& Rees(1978)]{Blandford1978}
Blandford, R.~D., \& Rees, M.~J. 1978, in BL Lac Objects, 
ed. A.~M. Wolfe (Pittsburgh, PA: Univ. Pittsburgh Press), 328

\bibitem[Caproni et al.(2017)]{Caproni2017}
Caproni, A., Abraham, Z., Motter, J.~C., \& Monteiro, H. 2017,
ApJ, 851, L39

\bibitem[Costamante \& Ghisellini(2002)]{Costamante2002}
Costamante, L., \& Ghisellini, G. 2002,
A\&A, 384, 56

\bibitem[Dash et al.(2026)]{Dash2026}
Dash, S., et al. 2026, 
MNRAS, in press.

\bibitem[Dermer(1995)]{Dermer1995}
Dermer, C.~D. 1995,
ApJ, 446, L63

\bibitem[Dinesh et al.(2025)]{Dinesh2025}
Dinesh, A., Dom{\'\i}nguez, A., Paliya, V., et al. 2025,
A\&A, 703, A162

\bibitem[Ghisellini et al.(2005)]{Ghisellini2005}
Ghisellini, G., Tavecchio, F., \& Chiaberge, M. 2005,
A\&A, 432, 401

\bibitem[Harutyunyan et al.(2026)]{Harutyunyan2026}
Harutyunyan, G., Sahakyan, N., B{\'e}gu{\'e}, D., \& Khachatryan, M. 2026,
MNRAS, 546, 1

\bibitem[Katarzy{\'n}ski et al.(2008)]{Katarzynski2008}
Katarzy{\'n}ski, K., Lenain, J.-P., Zech, A., Boisson, C., \& Sol, H. 2008,
MNRAS, 390, 371

\bibitem[Kirk et al.(1998)]{Kirk1998}
Kirk, J.~G., Rieger, F.~M., \& Mastichiadis, A. 1998,
A\&A, 333, 452

\bibitem[Kusunose \& Takahara(2008)]{Kusunose2008}
Kusunose, M., \& Takahara, F. 2008,
ApJ, 682, 784

\bibitem[Lind \& Blandford(1985)]{Lind1985}
Lind, K.~R., \& Blandford, R.~D. 1985,
ApJ, 295, 358

\bibitem[Madero \& Dom\'inguez (2026)]{Madero2026}
Madero, E., Dom\'inguez, A., 2026,
A\&A, 707, L18

\bibitem[Marscher et al.(2008)]{Marscher2008}
Marscher, A.~P., Jorstad, S.~G., D'Arcangelo, F.~D., et al. 2008,
Nature, 452, 966

\bibitem[Molteni et al.(1996)]{Molteni1996}
Molteni, D., Sponholz, H., \& Chakrabarti, S.~K. 1996,
ApJ, 457, 805

\bibitem[Pe{\~n}il et al.(2020)]{Penil2020}
Pe{\~n}il, P., Dom{\'\i}nguez, A., Buson, S., et al. 2020,
ApJ, 896, 134

\bibitem[Pe{\~n}il et al.(2024)]{Penil2024}
Pe{\~n}il, P., Westernacher-Schneider, J.~R., Ajello, M., et al. 2024,
MNRAS, 527, 10168

\bibitem[Raiteri et al.(2017)]{Raiteri2017}
Raiteri, C.~M., Villata, M., Acosta-Pulido, J.~A., et al. 2017,
Nature, 552, 374

\bibitem[Ren et al.(2023)]{Ren2023}
Ren, H.~X., Cerruti, M., \& Sahakyan, N. 2023,
A\&A, 672, A86

\bibitem[Rieger(2004)]{Rieger2004}
Rieger, F.~M. 2004,
ApJ, 615, L5

\bibitem[Rico et al.(2025)]{Rico2025}
Rico, A., Dom{\'\i}nguez, A., Pe{\~n}il, P., et al. 2025,
A\&A, 697, A35

\bibitem[Sandrinelli et al.(2014)]{Sandrinelli2014}
Sandrinelli, A., Covino, S., \& Treves, A. 2014,
ApJ, 793, L1

\bibitem[Sobacchi et al.(2017)]{Sobacchi2017}
Sobacchi, E., Sormani, M.~C., \& Stamerra, A. 2017,
MNRAS, 465, 161

\bibitem[Shukla et al.(2018)]{Shukla2018}
Shukla, A., Mannheim, K., Patel, S.~R., et al. 2018,
ApJ, 854, L26

\bibitem[Urry \& Padovani(1995)]{Urry1995}
Urry, C.~M., \& Padovani, P. 1995,
PASP, 107, 803

\bibitem[VanderPlas(2018)]{VanderPlas2018}
VanderPlas, J.~T. 2018,
ApJS, 236, 16

\bibitem[Vaughan et al.(2016)]{Vaughan2016}
Vaughan, S., Uttley, P., Markowitz, A.~G., et al. 2016,
MNRAS, 461, 3145

\bibitem[Villata et al.(2004)]{Villata2004}
Villata, M., Raiteri, C.~M., Aller, H.~D., et al. 2004,
A\&A, 424, 497

\bibitem[Zhang et al.(2014)]{Zhang2014}
Zhang, B.-K., Zhao, X.-Y., Wang, C.-X., et al. 2014,
Res. Astron. Astrophys., 14, 933

\bibitem[Zhang et al.(2017)]{Zhang2017}
Zhang, P.-f., Yan, D.-h., Liao, N.-h., et al. 2017,
ApJ, 835, 260

\end{thebibliography}
% - join the .bib files when you upload your source files
%%%%%%%%%%%%%%%%%%%%%%%%%%%%%%%%%%%%%%%%%%%%%%%%%%%%%%%%%%%%%%

\end{document}